\documentclass[cits]{PoS}

\usepackage{amssymb, amsthm, amsmath}
\usepackage{graphicx,graphics}
\usepackage{psfrag}
\usepackage{verbatim}
\usepackage{subfigure}
\usepackage{wrapfig}
\usepackage{amsfonts}

\newcommand{\field}[1]{\mathbb{#1}} 

\title{New phases of finite temperature gauge theory from an extended action}

\ShortTitle{New Phases of Finite Temperature Gauge Theory from an Extended Action}

\author{\speaker{Joyce Myers}
        \thanks{MCO and JCM gratefully acknowledge the support of the U.S. Dept of Energy.}\\
        Washington University in St. Louis\\
        E-mail: \email{jcmyers@wustl.edu}}

\author{Michael Ogilvie\\
        Washington University in St. Louis\\
        E-mail: \email{mco@wuphys.wustl.edu}}

\abstract{We study the behavior of the order parameter, the phase diagram, and the thermodynamics of exotic phases of finite temperature gauge theory. Lattice simulations were performed in $SU(3)$ and $SU(4)$ with an adjoint Polyakov loop term added to the standard Wilson action. In $SU(3)$, the pattern of $Z(3)$ symmetry breaking in the new phase is distinct from the pattern observed in the deconfined phase. In $SU(4)$, the $Z(4)$ symmetry is spontaneously broken down to $Z(2)$, representing a partially-confined phase. The existence of the new phases is confirmed in analytical calculations of the free energy based on high-temperature perturbation theory.}

\FullConference{The XXV International Symposium on Lattice Field Theory\\
		 July 30 - August 4 2007\\
		 Regensburg, Germany}

\begin{document}

\begin{section}{Background}

The deconfining phase transition of SU(N) ($N \geq 3$) gauge theories in $3+1$ dimensions is characterized by a low-temperature confined phase, where $Z(N)$ symmetry is unbroken and quarks and gluons are bound, and a high-temperature deconfined phase where $Z(N)$ symmetry is spontaneously broken and quarks and gluons are free \cite{Svetitsky:1982gs}. Simulations of pure gauge theories indicate that the transition between the confined and deconfined phases is first order for all $N \geq 3$. The global $Z(N)$ symmetry appears to always break completely in the deconfined phase, with no residual unbroken subgroup.



The confined phase of pure gauge theories is in a region of low temperature that cannot be accessed perturbatively. It is therefore useful to generalize the system to restore the confined phase in a region of high temperatures. We were motivated in part by Davies {\it et al.} who generalized the mechanism of color confinement in a monopole gas to four-dimensional supersymmetric gauge theories on $\field{R}^3 \times S^1$ \cite{Davies:1999uw}. They showed that monopole contributions to the superpotential led to an effective action with a $Z(N)$ symmetric minimum, corresponding to the confined phase, for all values of the $S^1$ circumference (naively analogous to temperature). Therefore it is reasonable to expect that the addition of a term to the pure gauge theory action which mimics the effects of monopoles would make the confined phase to accessible at all $\beta$. To this end we extended the Euclidean action of the pure $SU(N)$ gauge theory with a $Z(N)$ invariant term, the adjoint Polyakov loop:\vspace{-4mm}

\begin{equation}
-\int d^{3}x\, h_{A}\, Tr_{A}P(\vec{x})=-T\int_{0}^{\beta}dt\int d^{3}x\, h_{A}\, Tr_{A}P(\vec{x}).
\end{equation}

\noindent Here $P(\vec{x})$ is the Polyakov loop at the spatial location $\vec{x}$, given by the path ordered exponential of the temporal component of the gauge field.

A heuristic argument suggests that confinement is restored at high temperatures through variation of $h_A$. Consider minimization of the effective potential

\begin{equation}
V_{eff}=\sum_R v_{R}Tr_{R}P -Th_{A} Tr_{A}P.
\end{equation}

\noindent Because $Tr_{A}P=\left|Tr_{F}P\right|^{2}-1$, positive $h_{A}$ favors maximization of $Tr_A P$, which implies $\left|Tr_{F}P\right| > 0$. Thus $Z(N)$ symmetry is broken which suggests this region is in the deconfined phase. Negative $h_{A}$ favors minimization of $Tr_{A}P$, implying $Tr_{F}P = 0$, which defines the confined phase. Therefore for sufficiently negative $h_{A}$, the confined phase may be restored above the normal $h_A = 0$ deconfinement temperature. In the weak-coupling regime of high temperature we can calculate the effective potential, pressure, string tensions and 't Hooft loop surface tensions and examine their behavior in the restored confined phase resulting from the variation of $h_A$ (see also \cite{Ogilvie:2007la}).

\end{section} 
\begin{section}{$SU(3)$ Simulation Results}

Our simulations were performed in $SU(3)$ and $SU(4)$ by adding an adjoint Polyakov loop term to the standard lattice action:

\begin{equation}
S=S_{W}+\sum_{\vec{x}}\, H_{A}\, Tr_{A}P(\vec{x})\label{action}
\end{equation}

\noindent where $S_{W}$ is the Wilson action. The naive relationship between the variable lattice parameter $H_A$, and the parameter used in our analytical calculations $h_A$, is $H_{A} = h_{A}a^{3}$, but there is an additional unknown multiplicative renormalization factor.

\begin{wrapfigure}{l}{.5\textwidth}
  \begin{center}
    \includegraphics[width=7cm]{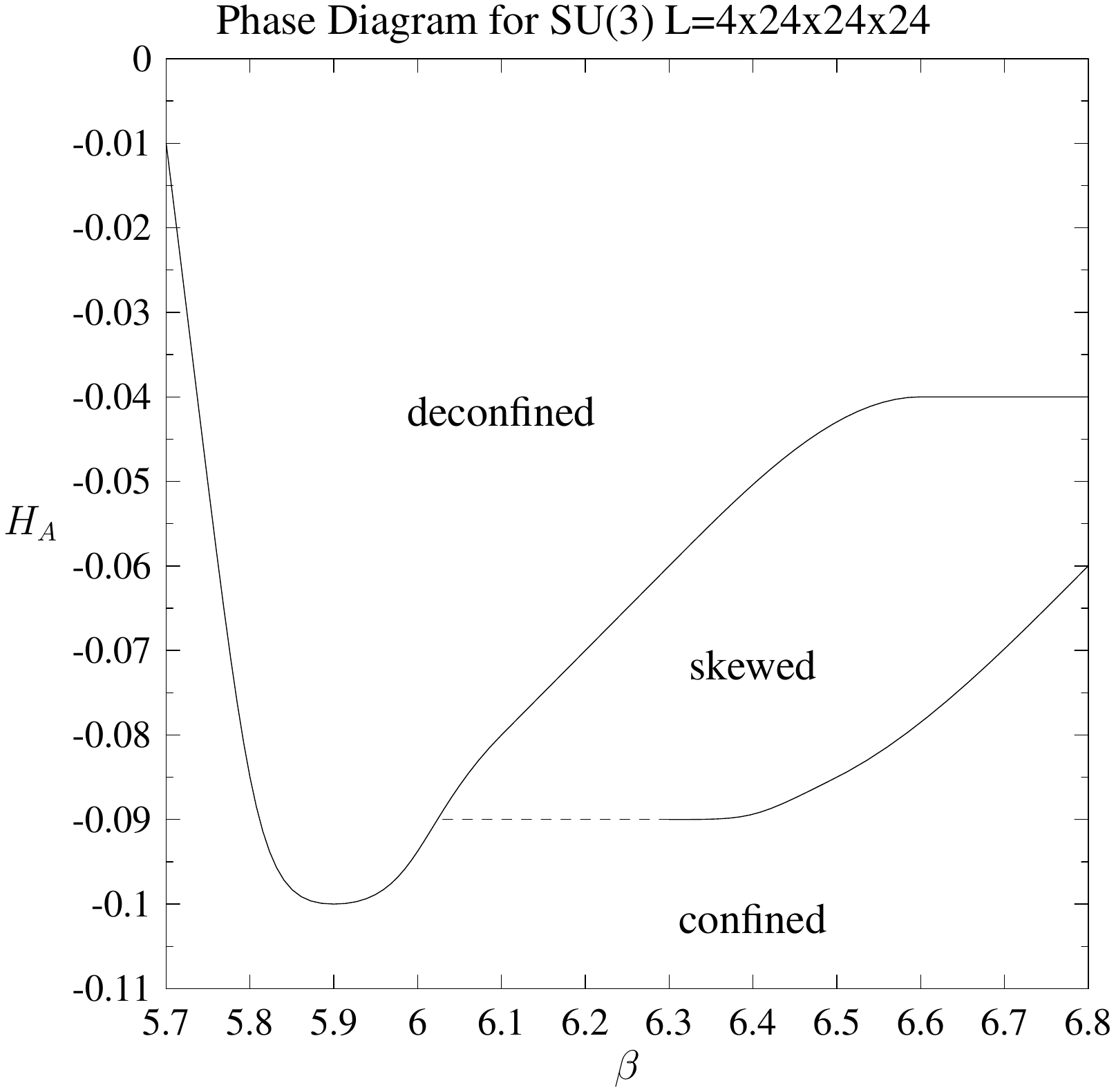}
  \end{center}
  \vspace{-5mm}
  \caption{Phase diagram in $SU(3)$ for an extended action}
  \vspace{5mm}
  \label{fig:1}
\end{wrapfigure}
\hspace{10mm}

Our simulation results for $SU(3)$ show that increasing positive $H_{A}$ decreases the deconfinement temperature as expected, and for sufficiently negative $H_A$ confinement is restored at high temperature. However, for negative $H_A$ there is an unexpected new phase which breaks $Z(3)$ symmetry in a peculiar way. 

Figure \ref{fig:1} shows the phase structure in the $\beta$ - $H_A$ parameter space of $SU(3)$ defined in terms of $\left\langle Tr_F P\right\rangle$, where projection to the nearest $Z(3)$ axis is understood. In the region of negative $H_A$ there are 3 distinct phases: the deconfined phase with $\left\langle Tr_{F}P\right\rangle > 0$, the confined phase with $\left\langle Tr_{F}P\right\rangle = 0$, and the new "skewed" phase with $\left\langle Tr_{F}P\right\rangle < 0$.

As shown in Figure \ref{fig:1}, decreasing $H_A$ at fixed $\beta > 6$, we encounter first the deconfined phase, then the skewed phase, then the confined phase. To obtain the locations of the phase transitions we use the histograms of the fundamental Polyakov loop in combination with plots of the adjoint Polyakov loop susceptibility. Figure \ref{fig:2} shows $SU(3)$ histograms of the fundamental Polyakov loop order parameter.

\begin{figure}[!h]
   \begin{center}
      \subfigure[deconfined]{\includegraphics[width=0.32\textwidth]{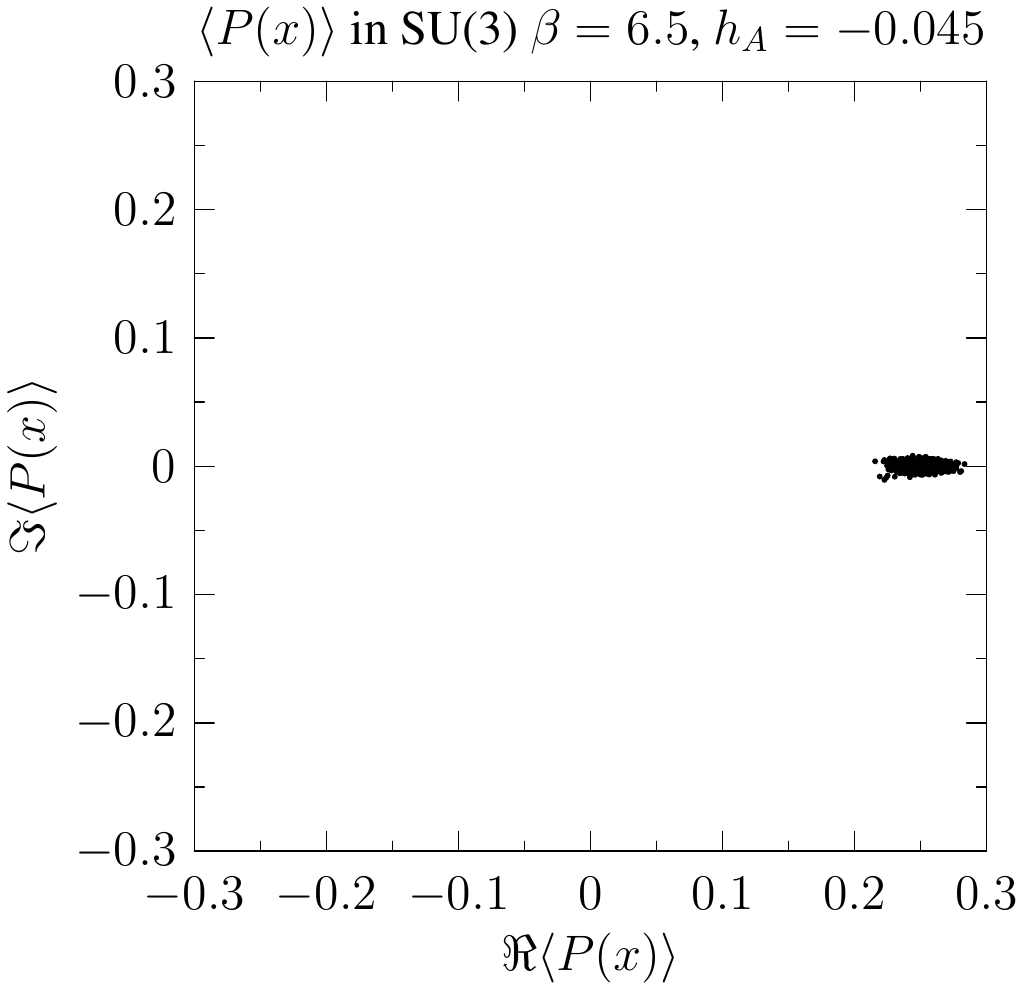}}
      \subfigure[skewed]{\includegraphics[width=0.32\textwidth]{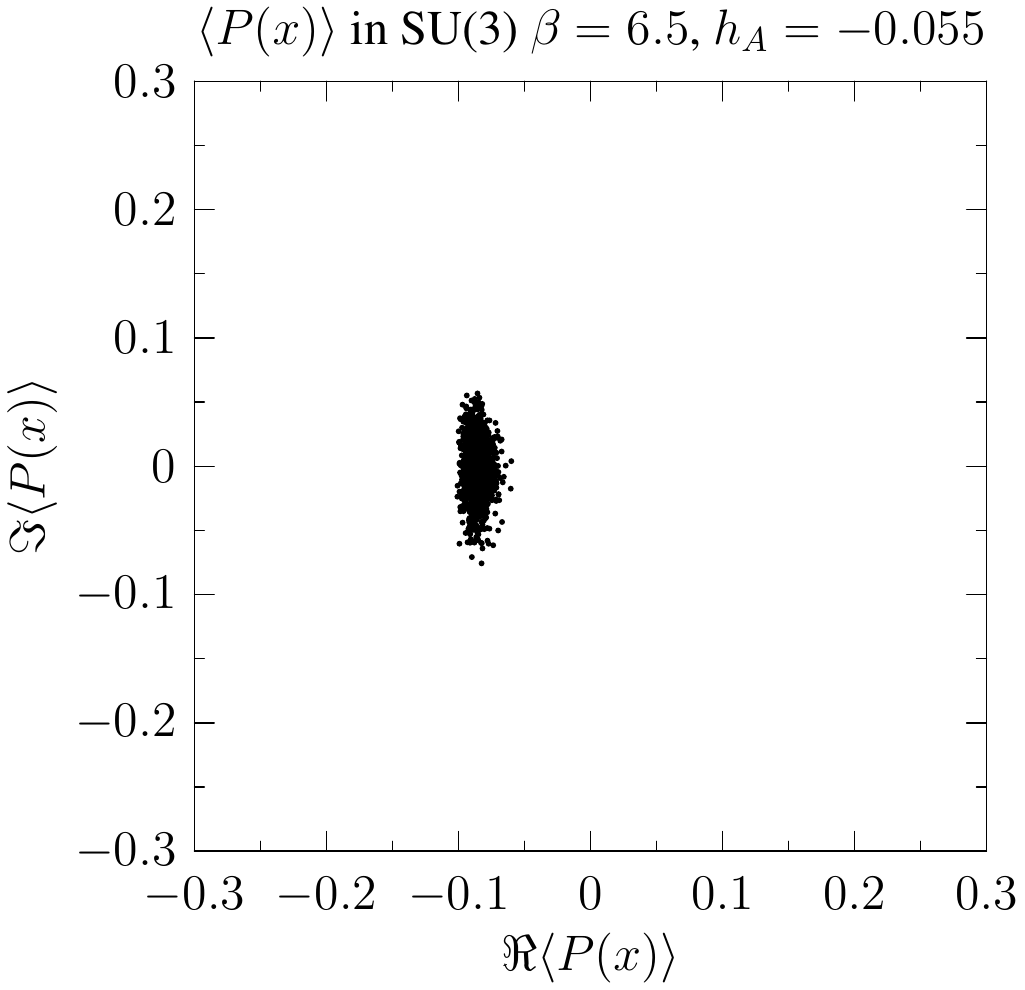}}
      \subfigure[skewed, with tunneling between phases]{\includegraphics[width=0.32\textwidth]{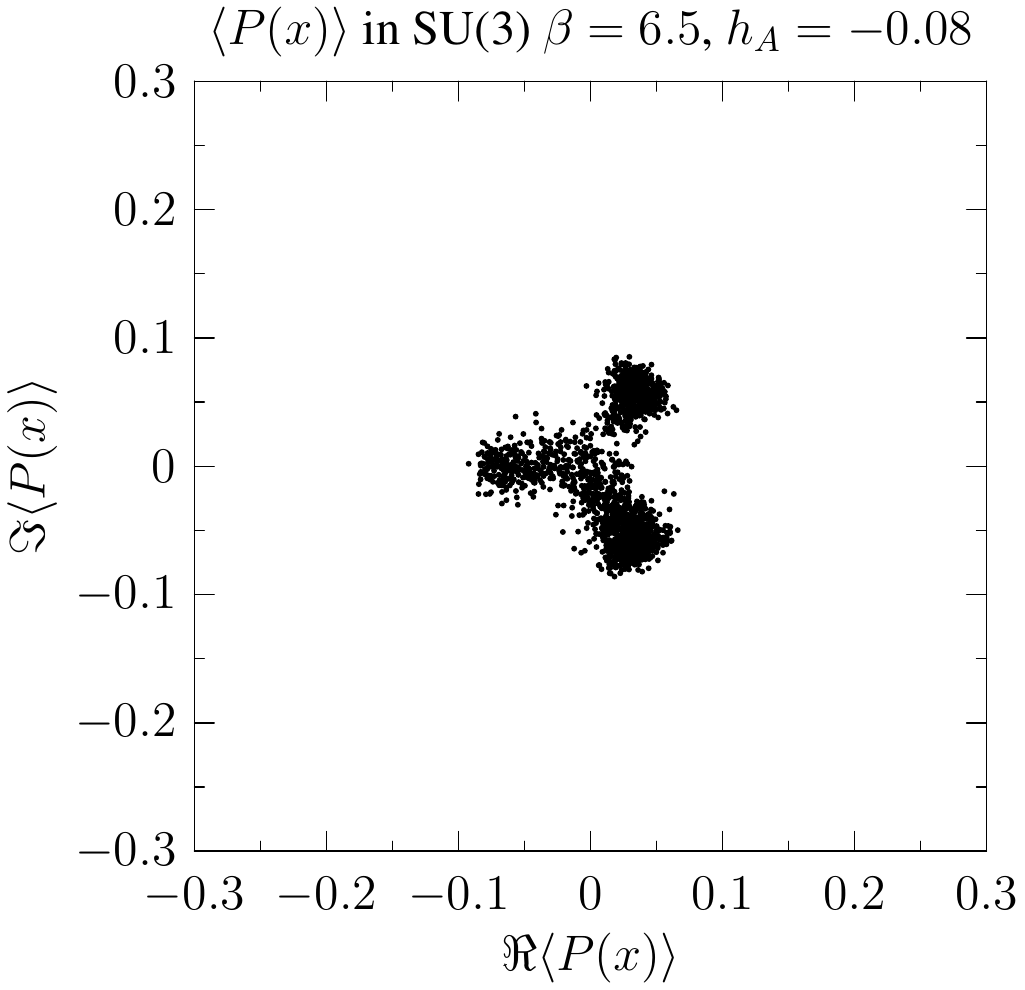}}
      \vspace{-2mm}
      \caption{$SU(3)$ Polyakov loop histograms}
      \label{fig:2}
   \end{center}
\end{figure}

\begin{figure}[!h]
  \hfill
  \begin{minipage}[t!]{.45\textwidth}
    \begin{center}  
      \includegraphics[width=0.85\textwidth]{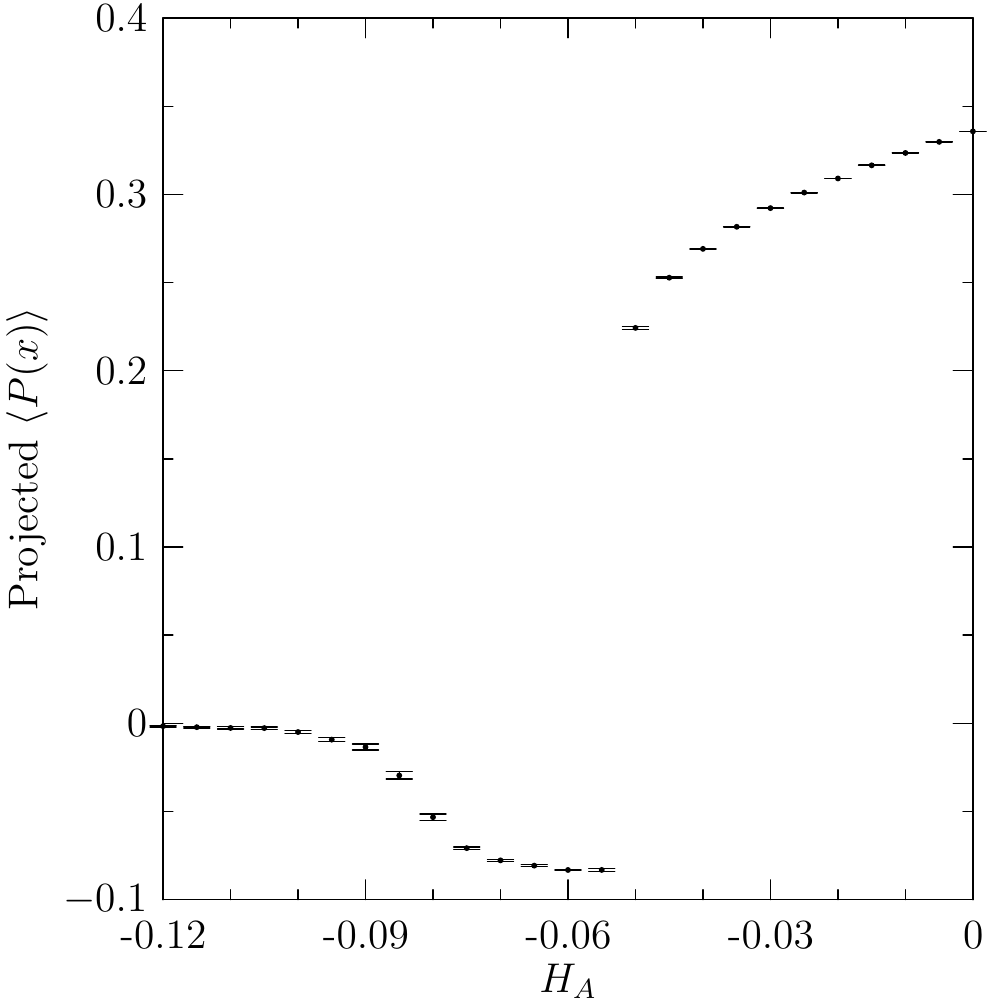}
      \vspace{-4mm}
      \caption{Projected $Tr_{F}P$}
      \label{fig:3}
    \end{center}
  \end{minipage}
  \begin{minipage}[t!]{.45\textwidth}
    \begin{center}  
 \includegraphics[width=0.98\textwidth]{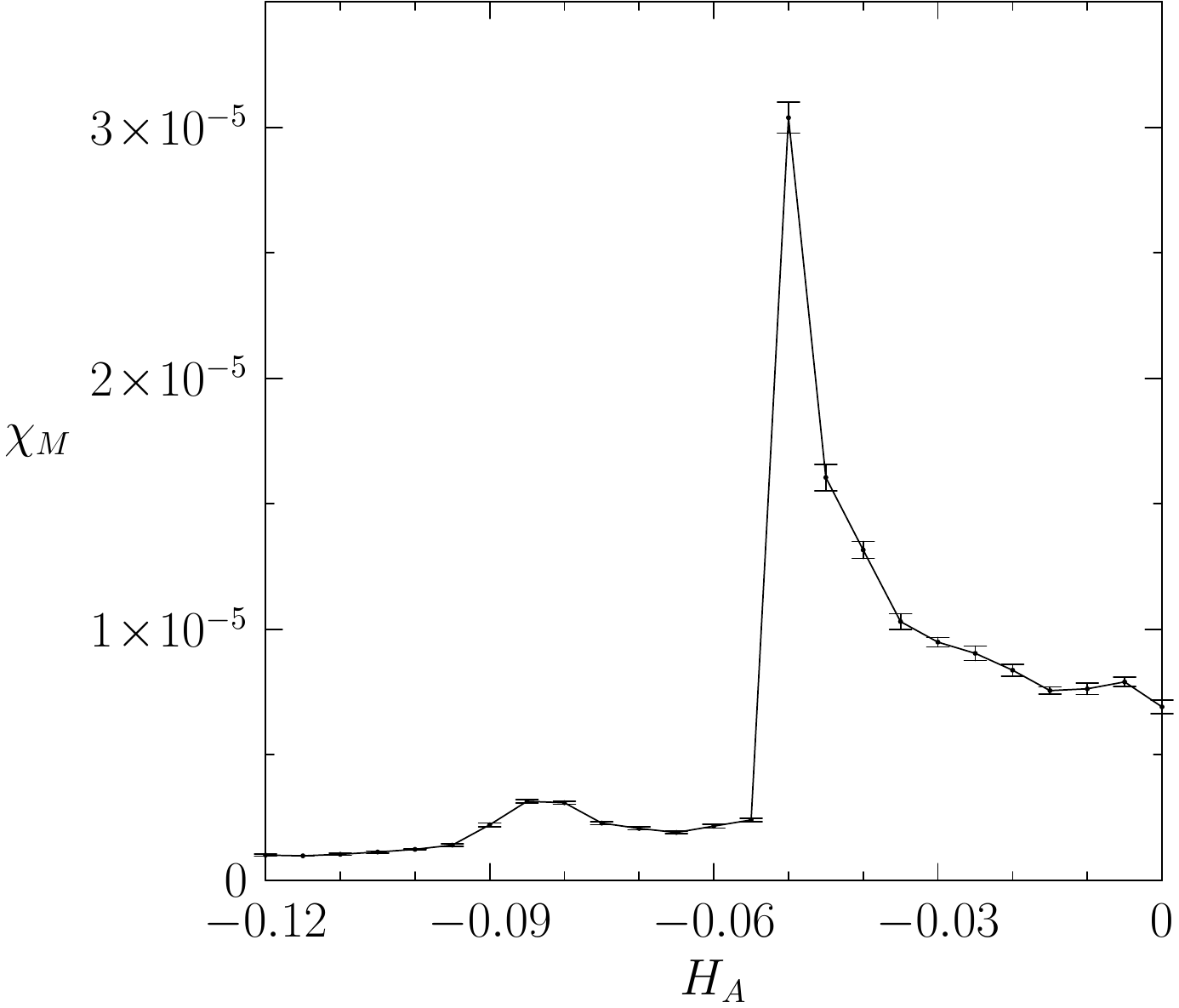}
      \vspace{-4mm}
      \caption{Adjoint susceptibility $\chi_{M}$}
      \label{fig:4}
    \end{center}
  \end{minipage}
  \hfill
\end{figure}

Figure \ref{fig:3} shows $\left\langle Tr_{F}P\right\rangle$ versus $H_A$. The presence of all three phases is clear. Figure \ref{fig:4} shows the adjoint Polyakov loop susceptibility. The obvious discontinuity in the order parameter shows that the transition between the deconfined and skewed phases is first-order in both graphs. The transition between the skewed phase and the confined phase is much weaker. It is likely to be first order as well, but this is not obvious due to the very small changes of the order parameter.


\end{section}

\begin{section}{$SU(3)$ Theory}

To confirm our lattice results we studied the thermodynamics of the system using an effective potential adapted from the one-loop free energy density first evaluated by Gross, Pisarski, and Yaffe \cite{Gross:1980br}, and by N. Weiss \cite{Weiss:1980rj}. Our modified expression is

\begin{equation}
V_{eff}=-2\frac{1}{2}Tr_{A}\int\frac{d^{3}k}{(2\pi)^{3}}\sum_{n}\ln[(\omega_{n}-A_{0})^{2}+k^{2}]-h_{A}T\, Tr_{A}P
\end{equation}

\noindent where the sum is over Matsubara frequencies $\omega_{n}=2\pi nT$. To locate the phases it is useful to write this as a function of the eigenvalues of the Polyakov loop:

\begin{equation}
\begin{aligned}
V_{eff} = &-2T^{4}\sum_{j,k=1}^{N}\left(1-\frac{1}{N}\delta_{jk}\right)\left[\frac{\pi^{2}}{90}-\frac{1}{48\pi^{2}}\left|\Delta\theta_{jk}\right|^{2}\left(2\pi-\left|\Delta\theta_{jk}\right|\right)^{2}\right]
\\
&-h_{A}T\left(\left|\sum_{j=1}^{N}e^{i\theta_{j}}\right|^{2}-1\right)
\end{aligned}
\end{equation}

\noindent In $SU(3)$, it is sufficient to consider $V_{eff}$ for the Polyakov loop projected onto the real axis, $P = diag [1,exp(i \phi),exp(-i \phi)]$. Figure \ref{fig:5} shows that the effective potential finds all 3 phases.

\begin{figure}[!h]
   \begin{center}  
      \subfigure[deconfined: $(Tr_F P)_{crit} = 3$]{\includegraphics[width=0.325\textwidth]{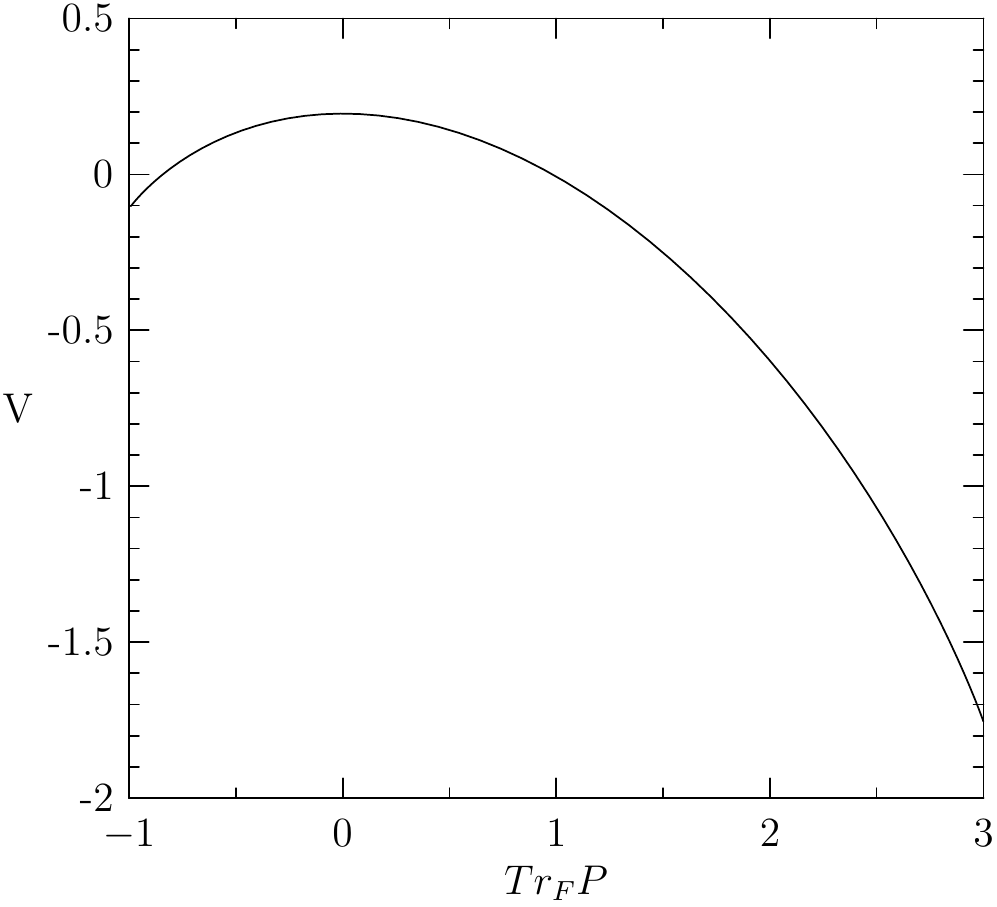}}
      \subfigure[skewed: $(Tr_F P)_{crit} = -1$]{\includegraphics[width=0.325\textwidth]{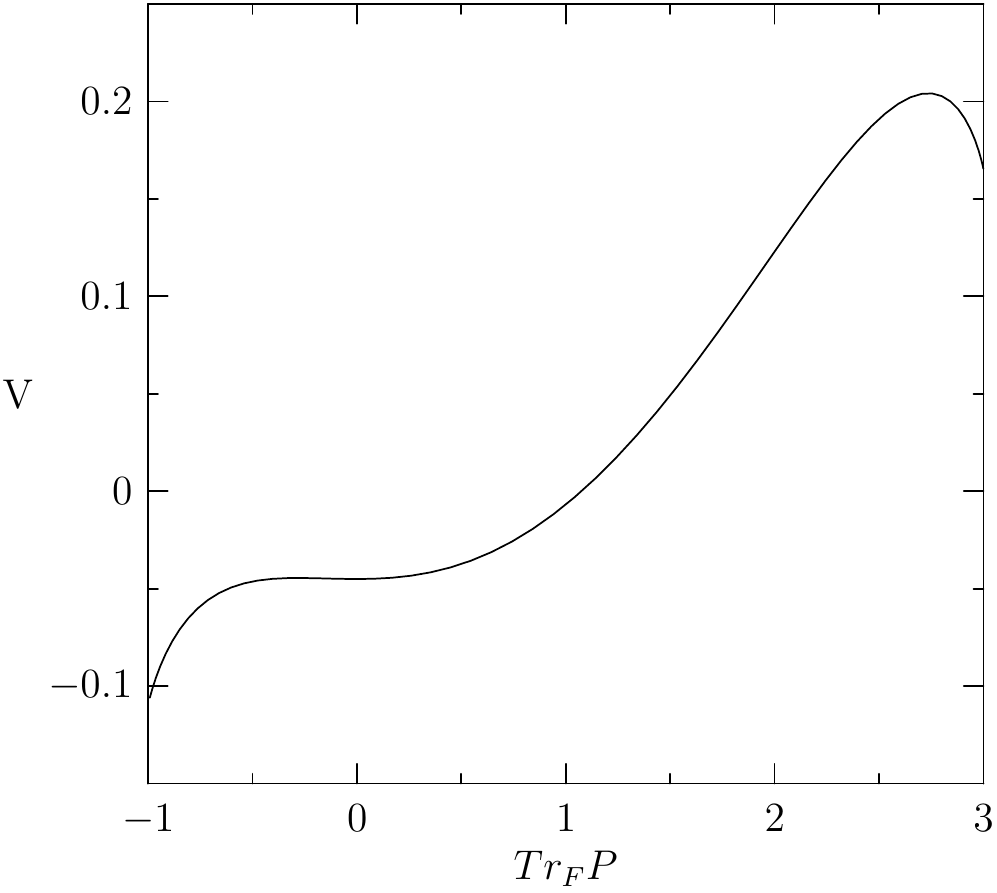}}
      \subfigure[confined: $(Tr_F P)_{crit} = 0$]{\includegraphics[width=0.325\textwidth]{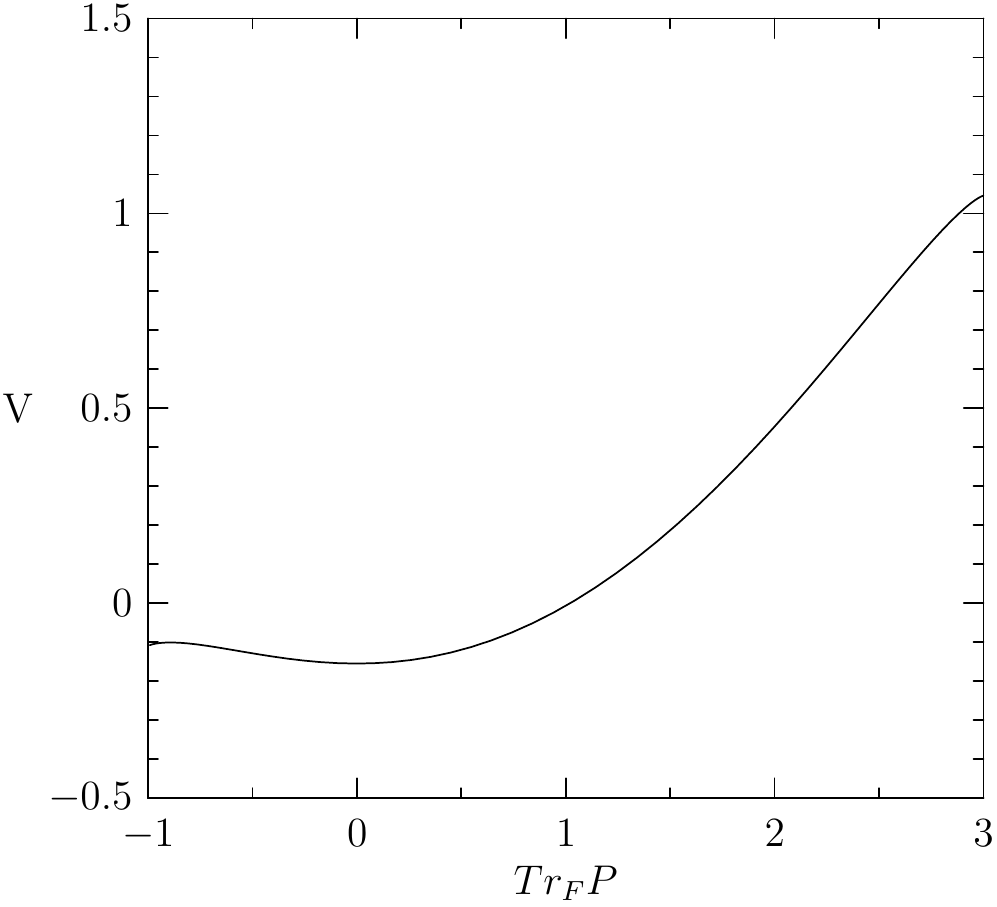}}
      \caption{Phases from calculation of $V_{eff}$ in $SU(3)$}
      \label{fig:5}
      \vspace{-4mm}
   \end{center}
\end{figure}

\end{section}
\begin{section}{Comparison of $SU(3)$ theory to simulation}

We have calculated the values of $\phi$ that minimize $V_{eff}$ in the three phases, then found the location of the phase transitions in terms of the dimensionless quantity $h_{A}/T^3$. The deconfined-skewed phase transition is located at $h_{A}/T^{3}=-\pi^{2}/48\simeq-0.206$. The skewed-confined phase transition is at $h_{A}/T^{3}=-5\pi^{2}/162\simeq-0.305$. The ratio of these values is similar to that from simulations.

\begin{wrapfigure}{l}{.5\textwidth}
  \begin{center}
  \vspace{-4mm}
     \includegraphics[width=70mm]{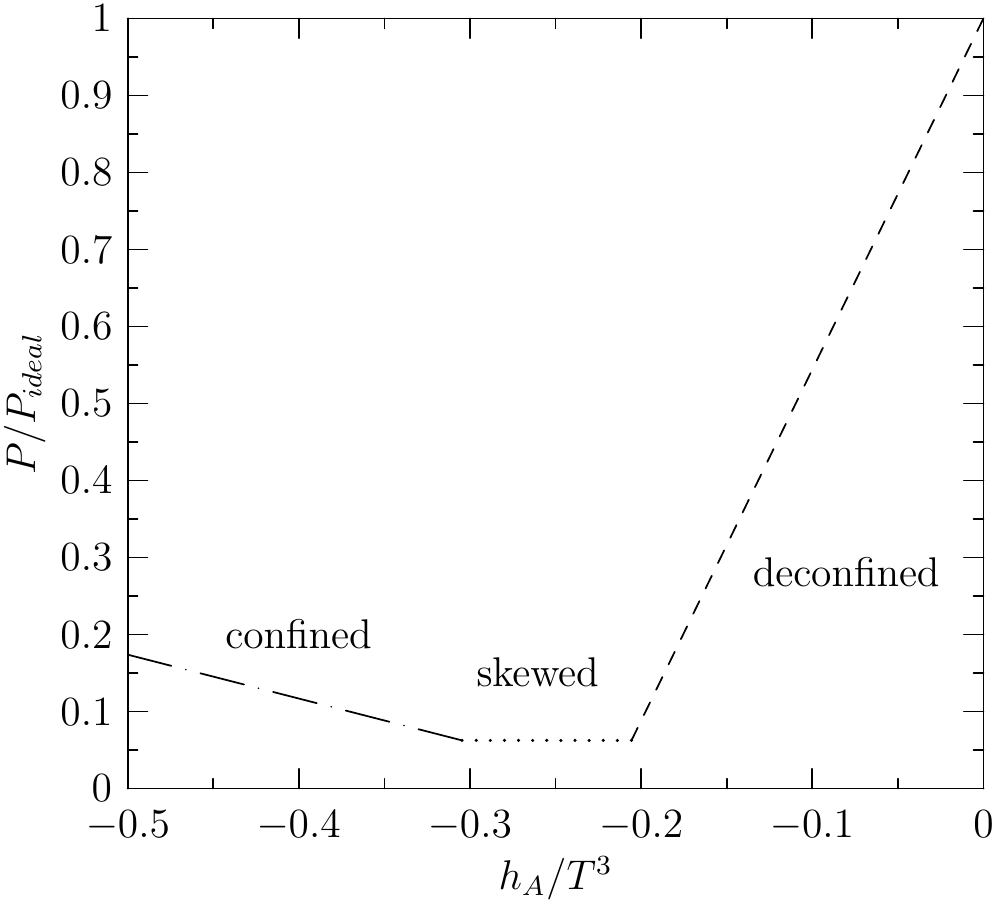}
  \end{center}
 \caption{Theoretical prediction for the pressure from $V_{eff}$ normalized by the black body value as a function of $h_A$.}
  \vspace{10mm}
  \label{fig:6}
\end{wrapfigure}

\hspace{2cm}

\vspace{-4mm}
We also compared values for the pressure determined from the effective potential to the pressure determined from simulations. Figure \ref{fig:6} shows the theoretical pressure from the effective potential. From simulations the pressure is calculated along a path of constant $\beta$ using

\begin{equation}
\frac{p_{2}}{T^{4}}-\frac{p_{1}}{T^{4}}=N_{t}^{3}\int_{1}^{2}dH_{A}\langle Tr_{A}P\rangle
\end{equation}.

\noindent Comparing $\Delta p/T^{4}$ across the deconfined and skewed phases we find for theory $\Delta p/T^{4} = \pi^{2}/6\simeq 1.64$ across the deconfined phase, and $\Delta p/T^{4} = 0$ across the skewed phase. In simulations $\Delta p/T^{4} = 1.64\pm0.03$ across the deconfined phase and $\Delta p/T^{4} = -0.18 \pm 0.07$ across the skewed phase.

\end{section}

\begin{section}{$SU(4)$ Simulation}

The case of $SU(4)$ is somewhat different. In simulations the new phase is partially-confining instead of skewed. Figure \ref{fig:7} shows the $SU(4)$ histograms of the fundamental Polyakov loop. The new phase again occurs for negative $H_A$. We first encounter the deconfined phase, then the partially-confined phase. Tunneling is observed as we continue decreasing $H_A$ in the partially confined phase. The fluctuations gradually reduce in size, but we are uncertain if there is a transition into the confined phase.

\begin{figure}[!h]
   \begin{center}  
      \subfigure[deconfined]{\includegraphics[width=0.32\textwidth]{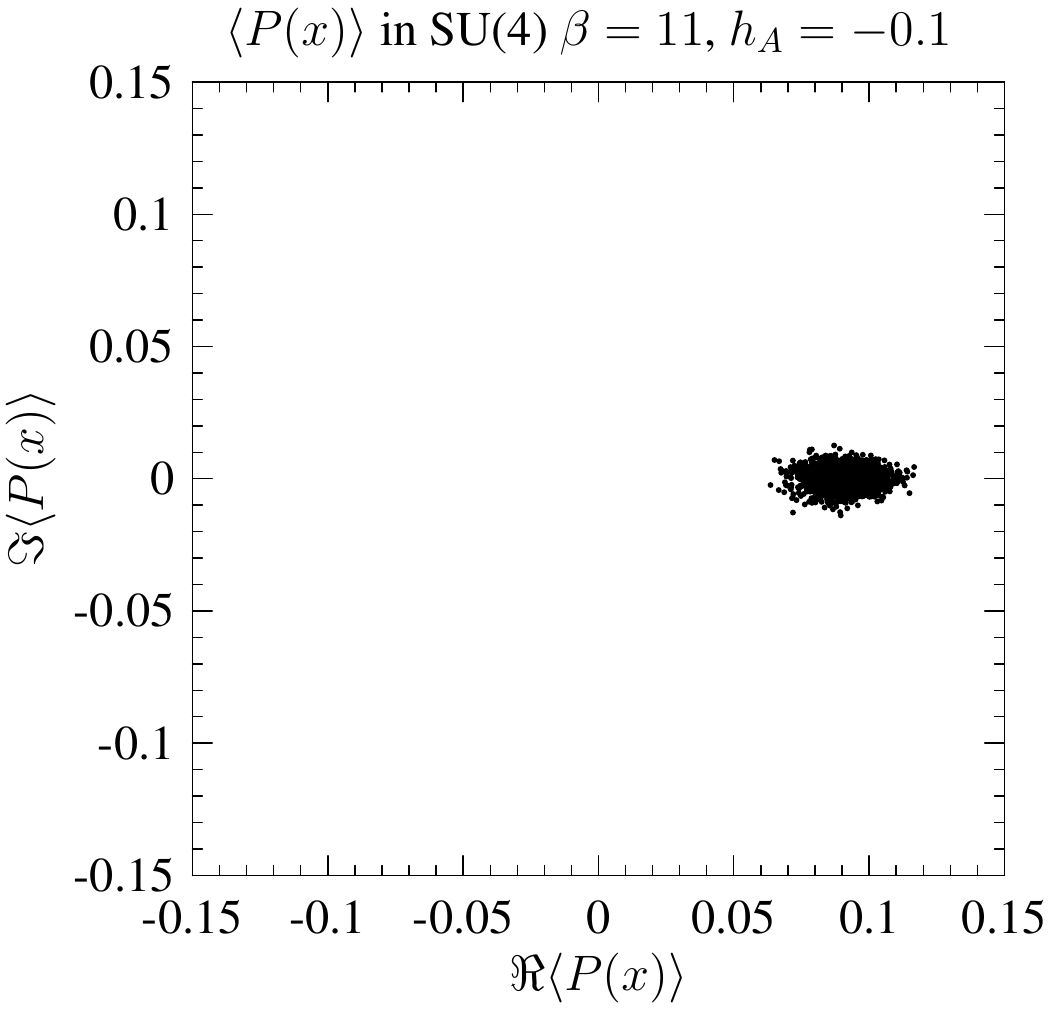}}
      \subfigure[partially confined]{\includegraphics[width=0.32\textwidth]{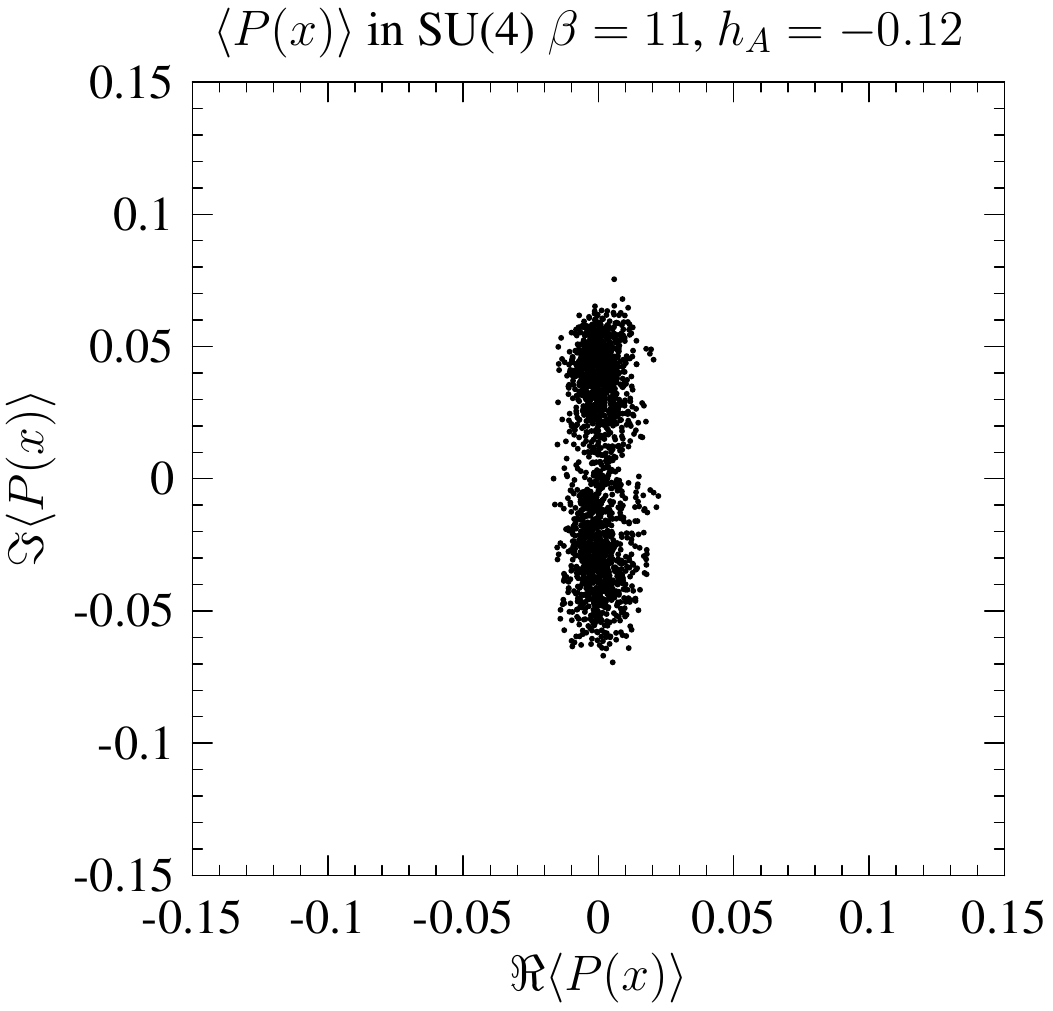}}
      \subfigure[partially confined, tunneling]{\includegraphics[width=0.32\textwidth]{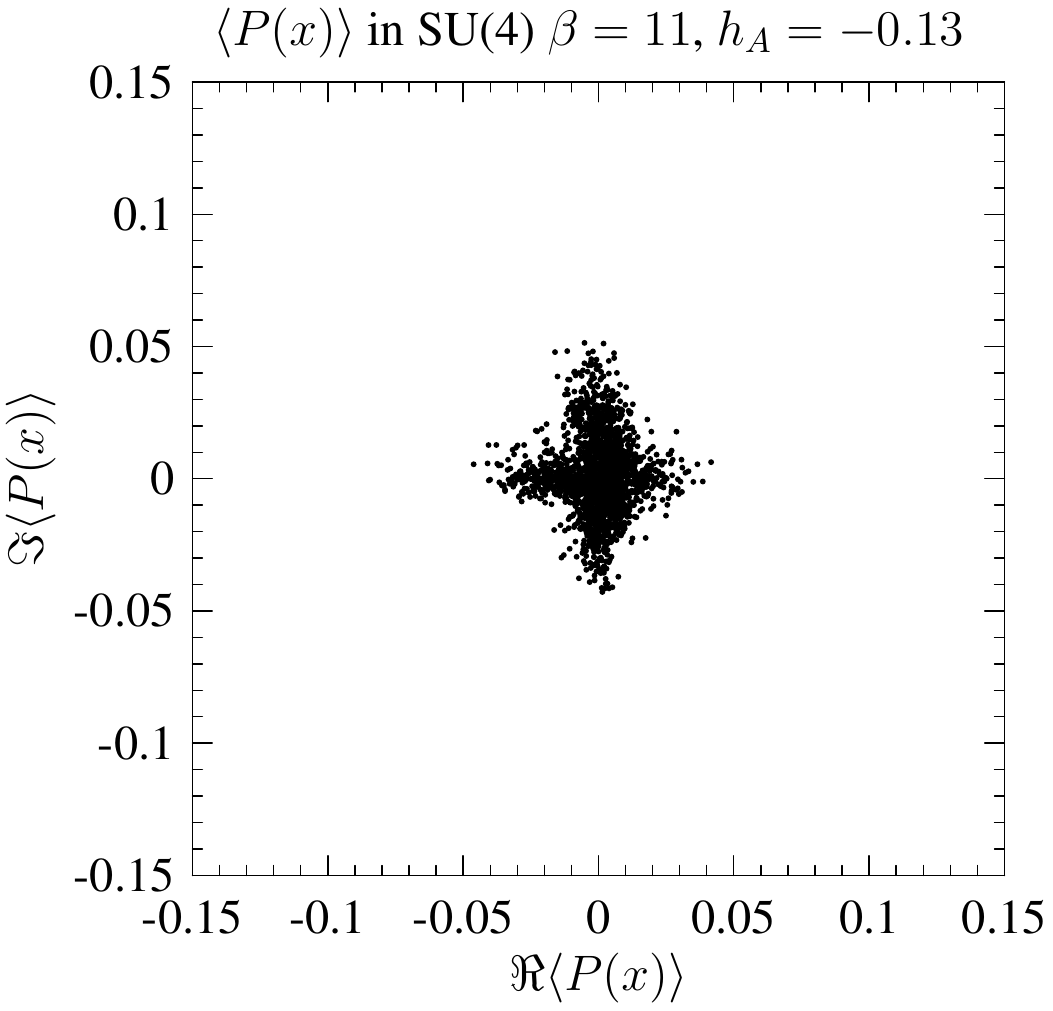}}
      \caption{$SU(4)$ Polyakov loop histograms}
      \label{fig:7}
   \end{center}
\end{figure}

In the new phase of $SU(4)$, global $Z(4)$ symmetry breaks spontaneously to $Z(2)$, a partially-confined phase. The residual $Z(2)$ symmetry ensures that $\left\langle Tr_{F}P\right\rangle = 0$, but that $\left\langle Tr_{R}P\right\rangle \neq 0$ for representations that transform trivially under $Z(2)$, so quarks are confined, but diquarks are not. The $Z(2)$ symmetry of the partially-confined phase is clear from the time history of variations of the real and imaginary parts of the Polyakov loop during a long run in which tunneling is observed, as shown in Figure \ref{fig:8}.

\begin{figure}[!h]
\includegraphics[width=.9\textwidth]{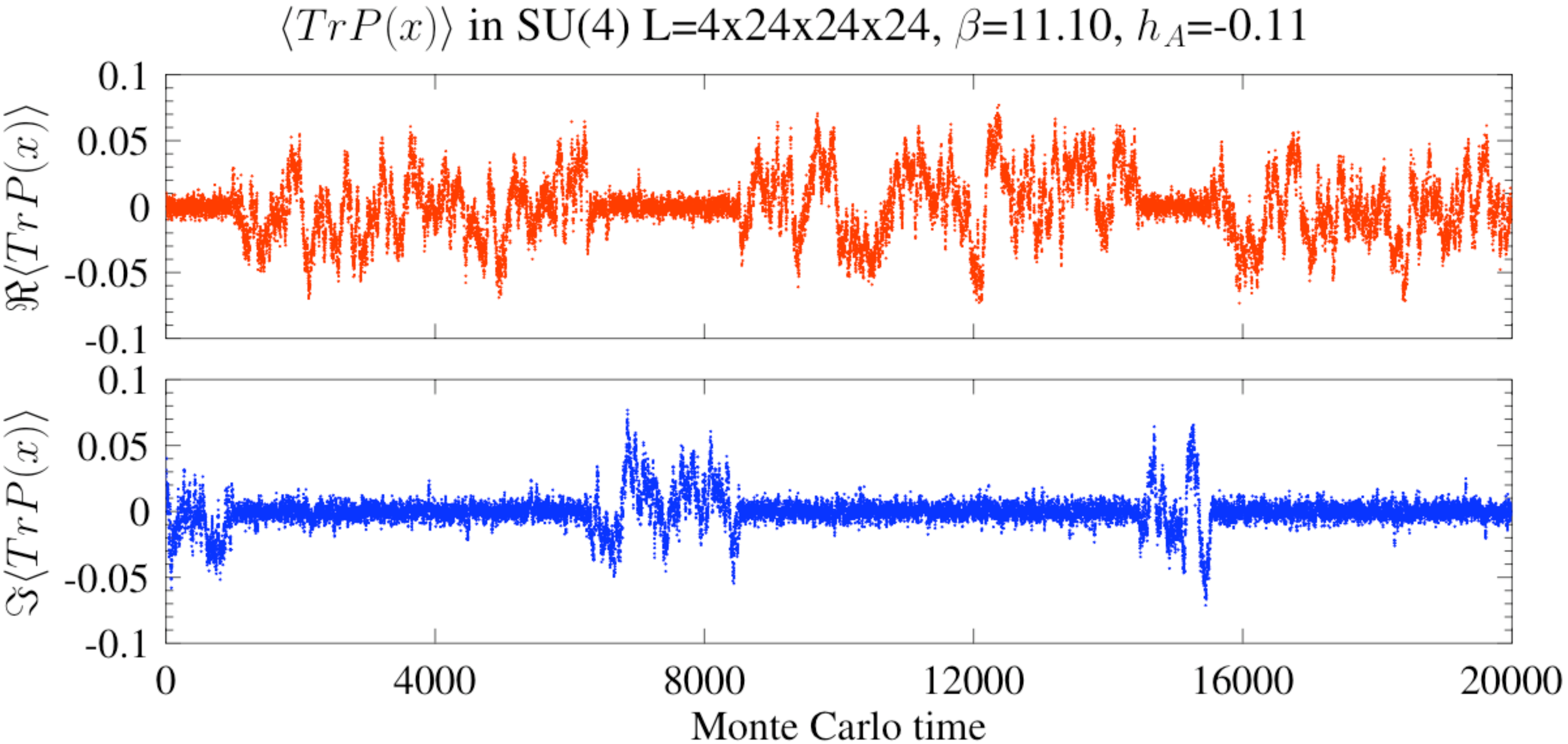}
\caption{Real and imaginary parts of $SU(4)$ Polyakov loop versus Monte Carlo time}
\label{fig:8}
\end{figure}

\end{section}


\begin{section}{$SU(4)$ Theory}

For our analytical calculations in $SU(4)$ we use again the one-loop effective potential to examine the possible occurrence of four different phases: the confined phase with full $Z(4)$ symmetry, the deconfined phase, a partially-confined $Z(2)$-invariant phase, and a skewed phase similar to that of $SU(3)$. However, only the deconfined phase and the $Z(2)$ phase are predicted by our simple theoretical model. A more complicated model with additional terms should reveal the confined phase \cite{Meisinger:2007jm}.


We compared the phase structure predicted by the one-loop effective potential with our simulation. $V_{eff}$ predicts a first-order transition between the deconfined and $Z(2)$-invariant
phases at $h_{A}/T^{3}=-\pi^{2}/48\simeq-0.205617$. This is in the same region of $h_A$ as in simulations. Across the deconfined phase, the theoretical value of $\Delta\left(p/T^{4}\right) = \pi^{2}/3\simeq 3.289$. In simulations $\Delta\left(p/T^{4}\right) = 2.21\pm0.07$


\end{section}


\begin{section}{Discussion and Conclusions}

We have considerable evidence, from lattice simulation and from theory, for the existence of new phases of finite temperature gauge theories in $SU(3)$ and $SU(4)$ when a $Z(N)$-invariant, adjoint Polyakov loop term is added to the gauge action. In $SU(3)$, confinement is restored at high temperatures, and the skewed phase was found.

It is interesting to note that Wozar {\it et al.} \cite{Wozar:2006fi}, in their study of $SU(3)$ spin models, observed a number of interesting new phases. One of these, which they refer to as the anti-centre phase, appears similar to our skewed phase. The anticenter phase resulted from an action of the form:

\begin{equation}
S_{eff} = \lambda_{F} S_{F} + \lambda_{15} S_{15}
\end{equation}

\noindent which includes a nearest neighbor coupling term in the $15$ representation instead of an adjoint potential term. We believe that these phases are related.

In the general case of $SU(N)$, there is good reason to expect a very rich phase structure may exist. For example, in $SU(6)$, we can consider partial breaking of $Z(6)$ to either $Z(2)$ or $Z(3)$. We have calculated the string tensions and 't Hooft loop surface tensions in the restored confined phase at high temperature \cite{Meisinger:2007jm}. These predictions can be checked in lattice simulations.

\end{section}

\end{document}